\documentclass[final,5p,times,fleqn]{elsarticle}

\usepackage{graphicx}
\usepackage{amssymb}
\usepackage{mathrsfs}
\usepackage{amsmath}

\usepackage{amsbsy}
\usepackage{mathrsfs}
\usepackage{amsfonts}
\usepackage{lineno,hyperref}
\usepackage{epsf,color,colordvi,pifont}

\journal{Physics Letters B}

\newcommand{\xim}{\xi}
\newcommand{\cep}{\chi}
\newcommand{\dynph}{\varphi}
\newcommand{\DPtoSP}{\varrho}  
\newcommand{\DPtoSUM}{\varrho} 
\newcommand{\ppden}{\eta} 

\begin{document}

\begin{frontmatter}

\title{Strong-Field Breit-Wheeler Pair Production\\in Two Consecutive Laser Pulses with Variable Time Delay}

\author{Martin J. A. Jansen and Carsten M\"uller}
\address{Institut f\"ur Theoretische Physik I, Heinrich-Heine-Universit\"at D\"usseldorf, Universit\"atsstr. 1, 40225 D\"usseldorf, Germany}

\begin{abstract}
Photoproduction of electron-positron pairs by the strong-field Breit-Wheeler process in an intense laser field is studied. The laser field is assumed to consist of two consecutive short pulses, with a variable time delay in between. By numerical calculations within the framework of scalar quantum electrodynamics, we demonstrate that the time delay exerts a strong impact on the pair-creation probability. For the case when both pulses are identical, the effect is traced back to the relative quantum phase of the interfering $S$-matrix amplitudes and explained within a simplified analytical model. Conversely, when the two laser pulses differ from each other, the pair-creation probability depends not only on the time delay but, in general, also on the temporal order of the pulses. 
\end{abstract}

\begin{keyword}
Electron-positron pair production\sep  short laser pulses \sep quantum interference
\MSC[2010] 00-01\sep  99-00
\end{keyword}

\end{frontmatter}

\section{Introduction}

The generation of matter-antimatter particle pairs from the electromagnetic energy of photons belongs to the most striking predictions of quantum electrodynamics (QED). It can be realized through the strong-field Breit-Wheeler (SFBW) reaction,
\begin{eqnarray}
\label{BW}
\omega_\gamma + n \omega \to e^+e^-\ ,
\end{eqnarray}
where a high-energy gamma ray of frequency $\omega_\gamma$ collides with a high-intensity laser field from which $n$ photons of frequency $\omega$ are absorbed to overcome the pair creation threshold. Experimental evidence for SFBW pair production was found in highly relativistic electron-laser collisions at SLAC \cite{Burke1997}. In the foreseeable future, further studies of the process are planned at high-intensity laser facilities such as the Extreme-Light Infra\-structure \cite{ELI}, the Exawatt Center for Extreme Light Studies \cite{XCELS}, the Diocles Petawatt Laser \cite{Diocles} or the HIBEF project \cite{HIBEF}. These campaigns are going to cover large areas of the parameter space for SFBW which have not been probed yet. As an alternative experimental approach to the Breit-Wheeler process, the usage of a thermal photon target has been proposed \cite{Pike2014,King2012}.

Since high laser intensities are generated in short pulses, theoreticians have started a few years ago to calculate pair production in laser fields of finite extent. With respect to the SFBW process, it was found that the broad frequency spectrum of a short pulse can strongly modify the energy and angular distributions of created particles \cite{Heinzl2010,Ipp2011,Krajewska2012BW,Fedotov2013,Hu2014,Meuren2016,Nousch2016,DiPiazza2016, Akal2016}. In particular, the carrier-envelope phase of a few-cycle pulse was shown to exert a characteristic impact \cite{Jansen2016a,Titov2016}. Besides, when several laser pulses follow each other, their partial contributions add up coherently, leading to a comb-like structure of emitted positrons \cite{Krajewska2014}. In certain laser parameter domains, the spectral broadness of a short pulse may also strongly affect the total pair creation probability due to subthreshold enhancement effects \cite{Titov2012, Nousch2012}. In laser-driven recollisions of a created electron and positron, even muon-antimuon pair production can result as a subsequent high-energy reaction \cite{Meuren2015b,Muller2008}.

The impact of finite pulses on pair production was also analyzed in other electromagnetic field configurations, such as time-dependent electric fields of finite duration or spatially localized electric and magnetic fields (see, e.g., \cite{Hebenstreit2009, Dumlu2011, Grobe2, Gies2016}). In particular, multiple-slit interference phenomena in the time domain were observed in sequences of electric-field pulses \cite{Akkermans2012, Li2014_bosons, Li2014}. Moreover, strong enhancement effects have been predicted when a rather weak, but fast oscillating field component is superimposed onto an intense, slowly varying field \cite{Schuetzhold2008, DiPiazza2009, Orthaber2011, Grobe1, Fey2012, Jansen2013, Augustin2014, Akal2014, Otto2015}. Systematic analyses to find the pulse shapes which optimize the pair yields were carried out \cite{Kohlfurst2013,Hebenstreit2014,Linder2015}. Also the creation of multiple pairs in electromagnetic fields of finite extension was addressed \cite{Grobe3, Wollert2016}.

\begin{figure}[b]
\centering
 \includegraphics[width=0.85\columnwidth]{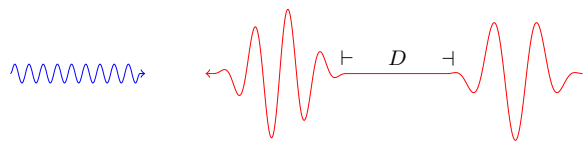}
\caption{Scheme of the field configuration: The gamma quantum (blue) collides with two consecutive short laser pulses (red) with a variable distance $D$.}
\label{Fig:Scheme}
\end{figure}

In this paper, we study SFBW pair production in a laser field which consists of two consecutive pulses, see Fig.~\ref{Fig:Scheme}. Our focus lies on effects arising from variations of the time delay  between both pulses. Two scenarios are considered: When both pulses are identical, the time delay is shown to strongly influence the energy spectrum of created particles and, remarkably, the total production probability, as well. 
Our numerical results are corroborated by a simplified analytical model.
When the pulses are different, the interesting question arises in addition whether their temporal sequence can affect the pair creation process. As we will show, in general the pulse order matters.

Gaussian units with $\hbar=c=1$ are employed throughout. The positron charge and mass are denoted by $e$ and $m$, respectively, and $\lambda_e = 1/m$ is the reduced Compton wavelength.

\section{Theoretical Framework}
The SFBW process is induced by the decay of a high-energy photon, which is described as one mode $\{{\bf k}_\gamma,\lambda_\gamma\}$ of a quantized radiation field $\hat{\mathcal{A}}^\mu$. Effectively, we employ the scattering potential 
\begin{equation}\label{AbsGamma}
 \mathcal{A}_\gamma^\mu = \langle 0| \hat{\mathcal{A}}^\mu | {\bf k}_\gamma \lambda_\gamma \rangle = \sqrt{\frac{2\pi}{V\omega_\gamma}} e^{-i k_\gamma\cdot x} \epsilon_\gamma^\mu\,,
\end{equation}
with the wave four-vector $k_\gamma^\mu =(\omega_\gamma,{\bf k}_\gamma)$ 
and a real polarization vector $\epsilon_\gamma^\mu$ fulfilling $k_\gamma\cdot\epsilon_\gamma=0$ 
and being referenced by a mode index $\lambda_\gamma$.
We use similar notation and conventions as in \cite{Jansen2016b}.

The two consecutive laser pulses are described classically by means of their combined vector potential
\begin{equation}
 \mathcal{A}^\mu = \mathcal{A}_1^\mu + \mathcal{A}_2^\mu\,,
\end{equation}
where each of the single pulses is of the form
$\mathcal{A}^\mu_j = \mathcal{A}^\mu_j (\phi_j) =  a_j f_j(\phi_j-\delta_j) \mathcal{X}_{[0,2\pi]}(\phi_j-\delta_j) \epsilon^\mu_j$,
with the amplitude parameter $a_j$ and phase variable
$\phi_j = k_j\cdot x$ for $j=1,2$. 
The actual shape is determined by the shape functions $f_j$ and the characteristic function $\mathcal{X}_{[0,2\pi]}(\phi)$ which is unity for $0\leq\phi\leq2\pi$ and zero otherwise.
The wave four-vectors $k_1$ and $k_2$ fulfill 
$k_1\cdot k_2 = 0$ and
$\epsilon_j \cdot k_j=0$, 
and $\epsilon_j^\mu$ is a real polarization four-vector.
The phase-shift parameters $\delta_j\geq0$ are chosen such that the pulses are strictly separated.
The particle states in the combined laser field $\mathcal{A}$ can be described by Gordon-Volkov solutions $\Psi_{p_\pm}$ (see, e.g., Eq. (5) in \cite{Jansen2016b}).
For calculational simplicity, the high-energy photon is assumed to collide head-on with the laser pulses.

Our calculations are performed within scalar QED, disregarding the electron and positron spin. This simplification helps us to render the main physical content of our study more transparent. Note that, in general, there can be significant differences between the creation of Klein-Gordon versus Dirac pairs, in particular on the basis of fully differential production probabilities. However, in terms of total probabilities, these differences diminish and reduce to an overall factor of about 3-5 for SFBW pair production in short laser pulses of moderate intensity \cite{Jansen2016b}. Also in the strong-field limit, the production rates of scalar and fermion pairs are known to coincide with each other, up to on overall prefactor \cite{Selym2013}. In the present paper, we shall mostly consider total production probabilities in double pulses, which are set into relation with the corresponding probabilities in single pulses. The basic influence from the double-pulse structure of the laser field can, thus, be expected to hold qualitatively for Dirac particles, as well. For further recent studies of strong-field pair creation within scalar QED, we refer to  \cite{Dumlu2011,Li2014_bosons,Fey2012,Jansen2013}.

The pair-creation amplitude is obtained from the $S$ matrix
\begin{equation}\label{Smatrix_double_start}
 \mathcal{S}_{p_+p_-}  = -i \int d^4x\,\Psi_{p_-}^* \, \mathcal{H}_\text{int}\, \Psi_{p_+}\,,
\end{equation}
with $\mathcal{H}_\text{int} = -ie\left(\mathcal{A}_\gamma \cdot \overset{\rightarrow}{\partial} - \overset{\leftarrow}{\partial} \cdot \mathcal{A}_\gamma\right) - 2e^2 \mathcal{A} \cdot \mathcal{A}_\gamma$ being the interaction Hamiltonian.
The $S$ matrix can be brought into the form
\begin{equation}\label{Smatrix_Double_Convenient}
 \mathcal{S}_{p_+p_-} =S_0 \int d^4x \, C \, e^{-iQ\cdot x-iH}\,,
\end{equation}
with
$S_0 = iem \sqrt{ \frac{\pi}{ 2 V^3 E_{p_+} E_{p_-} \omega_\gamma }}$
and the combined momentum vector 
$Q^\mu = k_\gamma^\mu - \left( p_+^\mu + p_-^\mu\right)$.
The reduced matrix element 
$C = C_0 + \sum_{j=1}^2 C_j$ 
contains the terms
$C_0 = \frac{p_--p_+}{m}\cdot \epsilon_\gamma$
and 
$C_j = \frac{2e\mathcal{A}_j(\phi_j)}{m}\cdot \epsilon_\gamma$.
The auxiliary function $H = H_1 + H_2$ can be decomposed into contributions from the individual pulses
\begin{equation}\label{HjDP}
 H_j = \int_0^{\phi_j} \sum_{l=1}^2 h_{l,j}f_j^l(\phi-\delta_j) \mathcal{X}_{[0,2\pi]}(\phi-\delta_j)\,d\phi\,, 
\end{equation}
with
$h_{1,j} = - ea_j \left[ \frac{\epsilon_j\cdot p_+}{k_j \cdot p_+} - \frac{\epsilon_j\cdot p_-}{k_j \cdot p_-}\right]$ and 
$h_{2,j} = -\frac{1}{2} e^2a_j^2 \left[ \frac{1}{k_j\cdot p_+} + \frac{1}{k_j\cdot p_-} \right]$.
For $\phi_j>\delta_j+2\pi$, the value of $H_j$ is constant and denoted as $H_j^\star$.

Switching to light-cone coordinates with $x^- = x^0-x^\parallel$ and $x^+ = \frac{1}{2} (x^0+x^\parallel)$, where $x^\parallel = {\bf k}_j \cdot {\bf x} / k_j^0$,
we obtain
\begin{equation}\label{Smatrix_Double_notrestricted}
 \mathcal{S}_{p_+p_-} = (2\pi)^3 S_0 \delta(Q^-)\delta^{(2)}({\bf Q}^\bot) \int dx^- \, C \, e^{-iQ^0x^--iH} \,.
\end{equation}
The remaining integral requires a regularization in analogy to the treatment presented in App. B of \cite{Krajewska2012Co}.
Effectively, we have to replace $C$ in Eq.~\eqref{Smatrix_Double_notrestricted} by the new matrix element $\tilde{C} = \tilde{C}_1 + \tilde{C}_2$ where each part $\tilde{C}_j = C_j-\frac{k_j^0}{Q^0} \frac{d H_j}{d \phi_j} \, C_0$ contains a characteristic function.
The pair-creation probability for unpolarized gamma quanta is obtained as
$\mathcal{P} = \frac{1}{2}\sum_{\lambda_\gamma} \int \frac{Vd^3p_+}{(2\pi)^3} \int \frac{Vd^3p_-}{(2\pi)^3} |\mathcal{S}_{p_+p_-} |^2$.

The pair-creation amplitude $\mathcal{S}_{p_+p_-}$ shall now be decomposed into contributions from the individual pulses, revealing the explicit dependence on the phase-shift parameters and thus the nature of the interaction.
To this end, we apply the formal substitution $x^- = (\Phi_j+\delta_j)/k_j^0$ to the integrals
\begin{equation}\label{EqIj}
 I_j = \int dx^- \tilde{C}_j e^{-iQ^0x^--i H_j}
\end{equation}
in order to shift the integration domains to $[0,2\pi]$. Accordingly, we can separate the dependence on $\delta_j$ and obtain
\begin{equation}\label{Ij_Fj}
 I_j = F_j \, e^{-iQ^0 \delta_j /k_j^0}
\end{equation}
with 
$ F_j = \frac{1}{k_j^0} \int_0^{2\pi} d\Phi_j \,\tilde{C}_j e^{-iQ^0 \Phi_j /k_j^0 - i H_j } $
being independent of $\delta_j$, since we can rewrite $H_j$ inside the integration domain of $F_j$ as 
$H_j = \sum_{l=1}^2 h_{l,j} \int_{0}^{\Phi_j} f_j^l(\tilde{\Phi}_j) \,d\tilde{\Phi}_j$, with $\Phi_j=\phi_j-\delta_j$. 
Similarly, $\tilde{C}_j$ is a function of $f_j(\Phi_j)$.
This way, the combined amplitude can be brought into the form 
\begin{equation}\label{SDP_F}
  \mathcal{S}_{p_+p_-} = (2\pi)^3 S_0\, \delta(Q^-)\delta^{(2)}({\bf Q}^\bot) \left( F_1 + F_2 \,e^{-i\dynph} \right)\,.
\end{equation}
Here, $\delta_1=0$ was chosen without loss of generality.
The contributions of the individual pulses to the pair-creation amplitude are given by $F_j$.
The dynamical phase
\begin{equation}\label{dynph_orig}
 \dynph = H_1^\star+Q^0 \Delta
\end{equation}
describes the phase propagation of the particles and the gamma quantum between the fronts of the pulses, which are separated by the distance $\Delta = \delta_2/k_2^0 = L_1 + D$. Here, $L_1$ is the length of the first pulse, and $D$ is the width of the gap between the pulses, see Fig.~\ref{Fig:Scheme}.
The free momenta are accounted for by the term $Q^0\Delta$, and the classical action of the charged particles in the first laser pulse is described by $H_1^\star$, which includes the Volkov phases.
The pair-creation process in the combined field is subject to quantum two-pathway interferences between the contributions $F_j$ of the individual pulses, with the interference phase being given by the dynamical phase $\dynph$.

\section{Numerical Results and Discussion}
Based on Eq.~\eqref{SDP_F}, we have calculated the probability of SFBW pair production in a double pulse.
Our examples are obtained for the pulse shape $f_j^\prime (\phi_j) = \sin^2(\phi_j/2)\sin(N_j \phi_j + \cep_j)$.
The characteristic function $\mathcal{X}_{[0,2\pi]}$ in the vector potential restricts the length $L_j = 2\pi N_j/\omega_j$ of the pulse and the number $N_j$ of field cycles.
The central pulse frequency is denoted by $\omega_j = N_j k_j^0$, and $\cep_j$ determines the carrier-envelope phase (CEP). We employ the field-strength parameter $\xim_j = \frac{ea_j}{m}\operatorname{max}_{\phi_j} |f_j(\phi_j)|$.
Throughout, the laser pulses share a common polarization direction $\epsilon_1 = \epsilon_2$, and the frequency of the gamma quantum is chosen as $\omega_\gamma = 1.01m$.
Note that the collision parameters discussed below always refer to a strongly Lorentz-boosted reference frame. They could be generated in a laboratory by using an optical laser of intensity $\sim\! 10^{18}$ W/cm$^2$ and a $\sim\!10$--$100$ GeV gamma quantum. The latter could be obtained from Compton backscattering off an ultrarelativistic electron beam, similarly to \cite{Burke1997}.

\subsection{Identical pulses}
First, we treat the case where the laser field consists of two identical pulses. The labels $j$ are dropped to simplify the notation, in particular $F_1=F_2 \equiv F$.
The differential probability in the combined field reads
\begin{equation}\label{DP_diff_prob}
 \frac{d^3 \mathcal{P}}{dp_+ d^2 \Omega_{p_+}} = \frac{e^2 m^2}{16\pi^2 \omega_\gamma} \sum_{\lambda_\gamma}  \frac{|{\bf p}_+|^2}{E_{p_+}(k_\gamma^- - p_+^-)} \,2 |F|^2  \left[1+\cos(\dynph)\right]\,,
\end{equation}
where we employ the notation $p_+=|{\bf p}_+|$.
In comparison with the case of a single pulse, the presence of the second pulse leads to a factor $2\left[1+\cos(\dynph)\right]$, which can, in the extreme cases, either enhance the probability by a factor of four, or suppress it completely. 
The two single pulses act like two identical slits in a double-slit experiment.
Similar interference phenomena in pulsed fields were discussed in \cite{Heinzl2010,Meuren2016,Krajewska2014, Hebenstreit2009,Dumlu2011,Akkermans2012,Li2014}.

\begin{figure}[ht]
\centering
 \includegraphics[width=0.8\columnwidth]{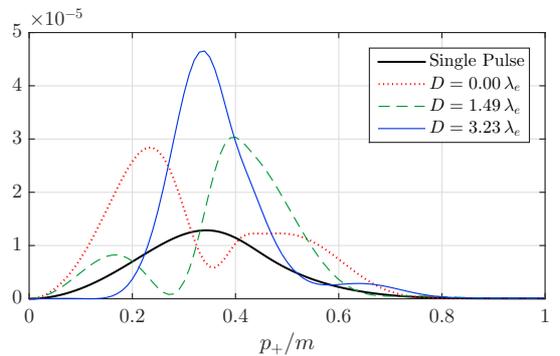}
\caption{Energy spectra $d\mathcal{P}/dp_+$ (in units of $1/m$) obtained for two identical pulses with different gap widths $D$ (see legend) and with $\xim=0.1$, $\omega=\omega_\gamma = 1.01m$, $N=4$ and $\cep=0$. For reference, the solid black line presents the energy spectrum obtained in a single pulse.}
\label{Fig:Ident_Mom_Spe}
\end{figure}

When the second pulse follows immediately after the first, such that $\Delta = L$ (i.e. $D=0$), the dynamical phase [see Eq.~\eqref{dynph_orig}] can be written as $-\dynph=E_L L$, with 
$E_L=-\left(Q^0+k^0\sum_{l=1}^2 h_{l} \langle f^l\rangle\right)$ 
being the laser-dressed energy which is required from the first pulse in order to produce the pair \cite{Jansen2016a}.
Here, we employ the pulse-length averages 
$\langle f^l\rangle = \frac{1}{2\pi }\int_0^{2\pi} f^l(\phi)\,d\phi$.
Since $E_L$ depends on ${\bf p}_+$, the dynamical phase $\dynph$ can completely change the angular distributions and energy spectra of the produced particles in comparison with a single pulse.
In Fig.~\ref{Fig:Ident_Mom_Spe}, we compare the energy spectra $d\mathcal{P}/dp_+$ for the case of a single pulse with $\xim=0.1$, $N=4$, $\omega=1.01m$ and $\cep=0$ (black solid line) and for the case when a second, identical pulse follows with $D=0$ (red dotted line). While the single-pulse spectrum has a rather simple structure with a broad maximum at $p_+\approx0.34m$, the double-pulse spectrum is more complex. It exhibits a maximum at $p_+\approx0.23m$ and a minimum at $p_+\approx0.36$, followed by a plateau region around $p_+\approx0.5m$.

These effects are all induced by the factor $2\left[1+\cos(\dynph)\right]$, which is averaged over the emission directions of the positron. For the present parameters, processes with $p_+\approx0.14m$ happen approximately in a c.m. system, inducing a common dynamical phase $\dynph$ for all angles. In fact, with $-\dynph\approx2\pi N = 8\pi$ since $E_L\approx\omega_\gamma = \omega$, the corresponding ratio between the double- and single-pulse probabilities is just slightly less than four, which is the maximum possible value. Especially for higher positron momenta, the required energy $E_L$ becomes angle dependent and the oscillating terms have a tendency to cancel. The plateau structure arises when the factor $\cos(\dynph)$ grows as a function of $p_+$ while the single-pulse probability drops.

When the second pulse arrives after a gap $D$, the dynamical phase receives the additional term $Q^0 D$, where $Q^0=-E_0$ can be recognized as the negative of the photon energy $E_0$ required to produce the pair without dressing effects. Accordingly, the dynamical phase [see Eq.~\eqref{dynph_orig}] can be expressed in the convenient form
\begin{equation}\label{Eq_DynPh}
 -\dynph = E_L L + E_0 D\,.
\end{equation}
Increasing the gap thus amplifies the momentum dependence of $\dynph$ and thereby enhances the modulating effects. 
For example, a gap width of $D=0.06L\approx1.5\lambda_e$, where $L\approx25\lambda_e$, induces a pronounced effect on the energy spectrum, which can be seen by comparing the green dashed line in Fig.~\ref{Fig:Ident_Mom_Spe} with the case of $D=0$ (red dotted line). 
In general, when $D$ is slightly increased, a given value of $\dynph$ can be achieved with a smaller energy. Accordingly, the locations of the extrema in the energy spectra are downshifted.
Further increasing the gap width to $D=0.13L\approx3.2\lambda_e$ (blue solid line) leads to a particular case, where the production of positrons in a relatively narrow energy range around $p_+\approx0.33m$ is enhanced by a factor of $\approx3.6$, while production of low-energy positrons is strongly suppressed.\footnote{For pair production in time-dependent electric field pulses, the impact of nonzero $D$ on the (fully) differential probabilities was revealed in \cite{Akkermans2012,Li2014,Kohlfurst2013}.}

\begin{figure}[ht]
\centering
 \includegraphics[width=1.00\columnwidth]{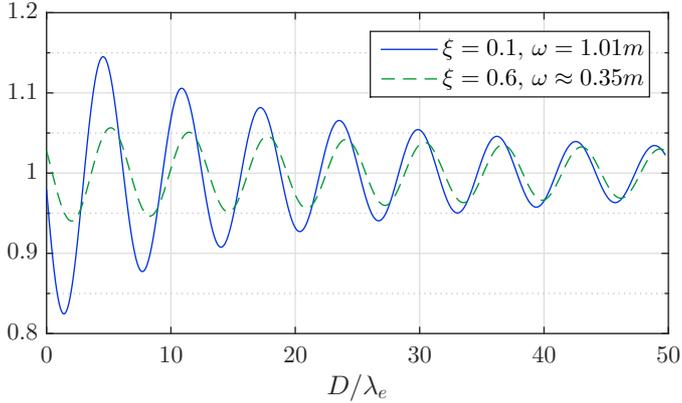}
\caption{Ratio $\DPtoSP$ between the total pair-production probabilities obtained in a double pulse comprising two identical pulses and in the single pulses, as function of the gap distance $D$. The pulse parameters for the solid blue line are the same as in Fig.~\ref{Fig:Ident_Mom_Spe}. The dashed green line presents the ratio $\DPtoSP$ for $\xim=0.6$, $\omega=0.3535m$, $N=4$ and $\cep=0$.}
\label{Fig:Ident_Dloop}
\end{figure}

Strong interference effects persist even in the total pair-pro\-duction probability. In Fig.~\ref{Fig:Ident_Dloop}, we present the ratio $\DPtoSP$ defined as the total probability in the double pulse divided by twice the total probability in a single pulse. As a function of the gap width $D$, the ratio exhibits a damped, oscillatory behavior.
The blue solid line is obtained for the same parameters as before. Starting with $\DPtoSP\approx0.98$ for $D=0$, the ratio arrives at its minimum value $\DPtoSP\approx0.82$ for $D\approx1.4\lambda_e$. The collimating effects found in the energy spectrum in Fig.~\ref{Fig:Ident_Mom_Spe} for $D\approx3.2\lambda_e$ are accompanied by only a slight enhancement of the total probability as compared to the single pulses, with $\DPtoSP\approx1.03$.
The maximum value $\DPtoSP\approx1.15$ is found for $D\approx4.5\lambda_e$, see Fig.~\ref{Fig:Ident_Dloop}.
For larger pulse distances $D$, the ratio $\DPtoSP(D)$ exhibits further oscillations with roughly constant period, but decreasing amplitude. 

This behavior can be understood as follows. The total pair-creation probability in the double pulse (comprising two identical pulses) is obtained as an integration of the form [see Eq.~\eqref{DP_diff_prob}]
\begin{equation}\label{Eq_PDP}
 \mathcal{P} = \int d^3 {\bf p}_+ \, \ppden({\bf p}_+) \, 2 \left[1 + \cos(\dynph) \right]\,,
\end{equation}
where $\ppden({\bf p}_+)$ denotes the single-pulse pair-production probability density. With regard to the dynamical phase $\dynph$ [see Eq.~\eqref{Eq_DynPh}], we recall that $E_L$ and $E_0$ both depend on ${\bf p}_+$.
In our previous examples with $\xim=0.1$, the main contributions to $\mathcal{P}$ stem from $|{\bf p}_+|\lesssim1.5m$.
When the momentum is varied, $\dynph$ induces oscillations, which become faster as the pulse distance grows. When $D$ is sufficiently large, they are much faster than any variation in $\ppden({\bf p}_+)$. Therefore, in the large-$D$ limit, the cosine term in Eq.~\eqref{Eq_PDP} can well be approximated by its average and thus vanishes. Hence, the ratio $\DPtoSP$ approaches the asymptotic value of~$1$, i.e., the combined probability is obtained as the sum of the contributions of the single pulses.

Regarding the fully differential probabilities in Eq.~\eqref{DP_diff_prob}, the cosine factor remains also for large values of $D$. Nevertheless, any detection process is based on an integration over a certain energy range or angular region. In the limit of large pulse distances, based on the same argument as before, the angular distributions and energy spectra of the detected particles thus approach the single-pulse patterns.

Further insights into the variation of $\DPtoSP$ with $D$ can be gained within a simplified model approach. It can be understood as the average of $1+\cos(\dynph)$ with respect to the probability density $\ppden$, i.e. $\DPtoSP = 1 + \langle \cos(\dynph)\rangle_\ppden$. Assuming that $E_L\approx E_0$, which is valid when dressing effects are small, the integration in Eq.~\eqref{Eq_PDP} can be expressed as an integral over the absorbed energies
\begin{equation}
 \mathcal{P} \approx \int d E_L \, \ppden(E_L) \, 2 \left[1 + \cos(E_L[L+D]) \right]\,.
\end{equation}
Here, $\ppden(E_L)$ is obtained as the two-dimensional integral of $\ppden({\bf p}_+)$ over those momenta with constant $E_L$. We denote the average absorbed energy by $\langle E_L \rangle$, and the width of the distribution $\ppden(E_L)$ by $\Delta E_L$. Assuming that $\ppden(E_L)$ can be described by a Gaussian distribution, we obtain
\begin{equation}\label{GaussMod}
 \DPtoSP \approx 1 + e^{-(\Delta E_L\left[L+D\right]/2)^2} \cos\left( \langle E_L\rangle [L+D]\right)\,.
\end{equation}
This expression illustrates in a semiquantitative manner how $\DPtoSP$ depends on the pulse distance. For large $D$, it asymptotically approaches the value of $1$, with the decay rate determined by the width $\Delta E_L$. Furthermore, $\DPtoSP$ oscillates in $D$, with a periodicity determined by the average absorbed photon energy $\langle E_L\rangle$.\footnote{We point out that the period length in $\DPtoSP(D)$ is not strictly constant and, for the present parameters, the decay rate is slower than suggested by the Gaussian model in Eq.~\eqref{GaussMod}. Besides, $\langle E_L\rangle$ and $\Delta E_L$ depend on the frequency spectrum of the pulse and the details of the pair-production process.}

For $\xim=0.1$ and $\omega \approx \omega_\gamma \approx m$, the process is mostly induced via absorbing one laser photon. The green dashed line in Fig.~\ref{Fig:Ident_Dloop} presents the ratio $\DPtoSP(D)$ for higher laser amplitude $\xim=0.6$ and smaller frequency $\omega \approx 0.35m$. 
For these parameters, at least three laser photons are absorbed.\footnote{The relevant numbers of absorbed photons can be estimated from the model approach to SFBW pair production developed in \cite{Jansen2016a}. Note that the dynamical phase is independent of the photon number, as long as the required energy $E_L$ remains constant.}
The interference effects persist, albeit with smaller amplitude, but roughly the same periodicity.
One reason for the smaller amplitude is the increased length $L\approx71\lambda_e$ of the pulses.
When $\xim$ is further increased, there is a general, though not strict, tendency that the oscillations in $\DPtoSP(D)$ decrease. For example, the maximum value of $\DPtoSP(D)$ is further reduced to about $1.015$ for $\xim=1.0$.
The general trend can be understood by noting that larger $\xim$ values upshift and broaden the positron energy distributions. As mentioned before, in the range of higher energies the dynamical phase oscillates more rapidly. This way, the integrated interference contributions are less pronounced, pushing the ratio $\DPtoSP$ towards $1$.

\subsection{Non-identical pulses}
\begin{figure}[ht]
\centering
 \includegraphics[width=1.00\columnwidth]{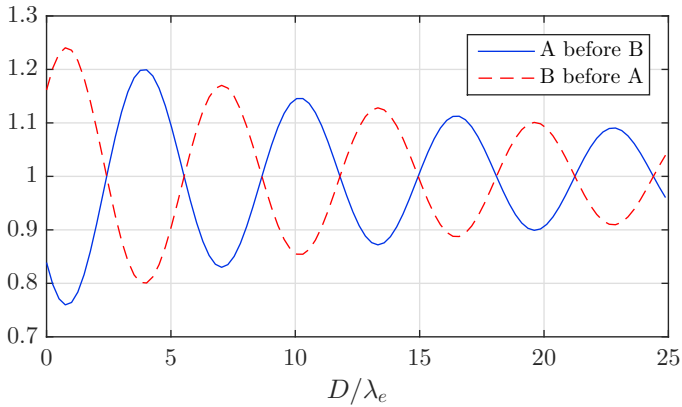}
\caption{Ratio $\DPtoSUM(D)$ between the total probability in the combined field of two different pulses A and B and the sum of the probabilities in the single pulses.
 The parameters are
$\xim_A = 0.1$, $N_A = 4$, $\omega_A = 1.01m$, $\cep_A = 0$, and
$\xim_B = 0.2$, $N_B = 3$, $\omega_B = 0.808m$, $\cep_B = \pi/2$.
The solid blue line depicts the case when pulse A arrives before pulse B, while the red dashed line shows the inverted order. 
}
\label{Fig:DivP_Sym}
\end{figure}

Now we regard two non-identical pulses A and B. Pulse A is the same pulse as employed in Fig.~\ref{Fig:Ident_Mom_Spe} with $\xim_A=0.1$, $\omega_A = 1.01m$, $N_A=4$ and $\cep_A=0$, while pulse B has higher field strength $\xim_B=0.2$, smaller frequency $\omega_B=0.808m$, only $N_B=3$ cycles and different CEP $\cep_B=\pi/2$. 
With these pulses being quite different, the question arises if the interference effects persist, and how they are affected by the temporal order of the pulses. 
We begin with the case when pulse A arrives first, and pulse B follows after a gap $D$. This configuration is depicted schematically in Fig.~\ref{Fig:Scheme}.
The blue solid line in Fig.~\ref{Fig:DivP_Sym} shows the ratio $\DPtoSUM(D)$ between the total probability obtained in the combined field and the sum of the probabilities in the single pulses A and B.
The latter are related as $\mathcal{P}_{B}\approx0.65\,\mathcal{P}_{A}$.

Inspecting Fig.~\ref{Fig:DivP_Sym}, we find that $\DPtoSUM(D)$ exhibits damped oscillations,
which closely resemble the overall behavior of $\DPtoSP(D)$ in the case of two identical pulses, see Fig.~\ref{Fig:Ident_Dloop}. In principle, the interferences are in both cases subject to a similar mechanism. 
However, when two non-identical pulses are employed, the interference terms are not only determined by the dynamical phase $\dynph$, but also by the complex values of the contributions $F_1\equiv F_A$ and $F_2\equiv F_B$ of the individual pulses, see Eq.~\eqref{SDP_F}.

When the order of the pulses is inverted, the corresponding ratio $\DPtoSUM(D)$ exhibits striking differences (red dashed line in Fig.~\ref{Fig:DivP_Sym}).
For example at $D\approx0.75\lambda_e$, we find $\DPtoSUM_{BA} \approx 1.24$ when B arrives before A, while $\DPtoSUM_{AB}\approx0.76$ for the other case. 
The qualitative dependence of $\DPtoSUM_{BA}$ on the pulse distance is very similar to the previous examples, but the entire structure appears to be inverted in comparison with $\DPtoSUM_{AB}$.
In fact, the total probabilities fulfill the relation
$ \frac{1}{2} \left[ \mathcal{P}_{AB}(D) +  \mathcal{P}_{BA}(D) \right] \approx  \mathcal{P}_{A} + \mathcal{P}_{B}$.

When we set $\cep_B=0$, keeping all other parameters fixed, we 
find instead that the total probability in the combined laser field is independent of the temporal order, i.e. $\mathcal{P}_{AB}(D) = \mathcal{P}_{BA}(D)$. Nevertheless, pronounced interference effects arise when $D$ is varied, and the corresponding ratio $\DPtoSUM(D)$ appears qualitatively similar to the red dashed line in Fig.~\ref{Fig:DivP_Sym}. Moreover, even the fully differential probability is invariant under the exchange of the pulses. 
This is a surprising result, given the high sensitivity of the process on the properties of the driving laser field.
The laser potential fulfills $\mathcal{A}_{AB}(\phi) = \mathcal{A}_{BA}(-\phi)$ in this case. 
A similar relation was found for pair production in time-dependent electric fields \cite{Li2014}. Note, however, that in our case not only the laser field but also the gamma quantum is required for the process [see Eq.~\eqref{AbsGamma}].
Conversely, for $\cep_A=\pi/2=\cep_B$, where $\mathcal{A}_{AB}(\phi) = -\mathcal{A}_{BA}(-\phi)$ holds, the fully differential probabilities depend on the ordering, while
the total probability is still invariant.

An intuitive insight into the complex influence of the CEPs is gained by the notion that, instead of two consecutive stages, the two pulses rather act like slits in a double-slit experiment, which are probed simultaneously by the (infinitely extended) gamma quantum. 
In this picture, the properties of the non-identical pulses translate to differently structured slits.
The CEPs affect the symmetry properties of each single pulse, which, in turn, jointly determine the global symmetry of the double-slit.

\begin{figure}[t]
\centering
 \includegraphics[width=1.00\columnwidth]{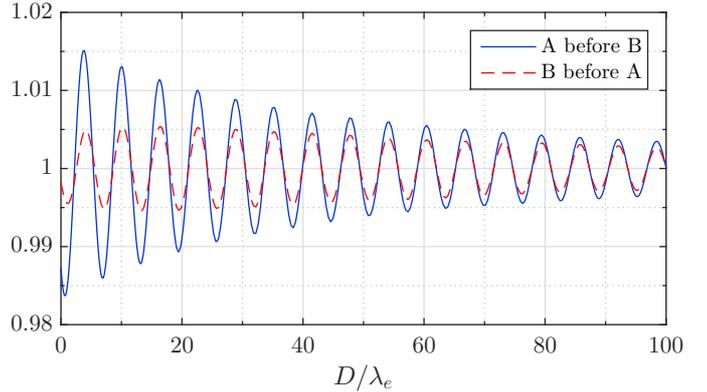}
\caption{Same as Fig.~4, but with pulse B chosen as $\xim_B=1.0$, $N_B=4$, $\omega_B=0.35m$, and, in particular, a different CEP of $\cep_B=\pi/4$.}
\label{Fig:DivP_ASym}
\end{figure}

Finally, we regard a situation where pulse A remains the same as in Fig.~\ref{Fig:DivP_Sym}, but pulse B has a substantially higher intensity parameter of $\xi_B = 1.0$, lying in the nonperturbative interaction domain. The other parameters are $\omega_B = 0.35m$, $N_B = 4$, and $\chi_B = \pi/4$. We depict the ratio $\DPtoSUM(D)$ for the two different configurations in Fig.~\ref{Fig:DivP_ASym}. Again, the total probability clearly depends on the temporal order of the pulses and exhibits characteristic oscillations as the pulse distance grows. In $\DPtoSUM_{AB}$, these oscillations are damped as before, whereas $\DPtoSUM_{BA}$ exhibits a region of almost constant oscillation amplitude for small values of $D$. This implies that, for certain parameter combinations, the qualitative appearance of $\DPtoSUM(D)$ may differ from the prediction of our simplified analytical model [see Eq.~\eqref{GaussMod}].
Overall, the amplitude of the oscillations is largely reduced as compared with Fig.~\ref{Fig:DivP_Sym}.
In addition to the rapidly oscillating quantum phases in $F_A+e^{-i\dynph}F_B$, the higher field strength of pulse B facilitates pair creation in a substantially wider range of energies (with different angular distributions), such that there is less possibility for interference with pulse A.\footnote{In principle, also the fact that the pair creation probabilities of the single pulses are quite different, with $\mathcal{P}_A \approx 0.28 \,\mathcal{P}_B$, may lead to reduced interference effects. We note, however, that almost the same oscillation amplitudes as in Fig.~\ref{Fig:DivP_ASym} result when the intensity of pulse A is increased to $\xi_A = 0.15$, which corresponds to $\mathcal{P}_A \approx 0.61\, \mathcal{P}_B$.}

The overall appearances of $\DPtoSUM_{AB}(D)$ and $\DPtoSUM_{BA}(D)$ become more similar when the CEPs are switched to $\chi_A = \pi/4$ and $\chi_B = 0$. Both curves exhibit damped oscillations within a common envelope then. 
The more similar behavior of the production probabilities might be related to the fact that, this time, the strong pulse B is symmetrically shaped (whereas in Fig.~\ref{Fig:DivP_ASym} this holds for the weak pulse A). In a laser pulse with $\xi\gtrsim 1$, pairs are mainly created in phase regions around the electric field maxima (see, for example, \cite{Meuren2016}). In the case, when such a pulse is asymmetric due to the impact of the CEP, the effective distance $D_{\rm eff}$ to the dominant pair-creation zone changes when this pulse preceeds before or follows after a symmetrically shaped weaker pulse. As a result, the influence of the pulse ordering can be enhanced.

For $\chi_A=\chi_B = 0$, the pair-production probabilities (total as well as fully differential) remain unchanged when the pulse order is switched. While the same property was also observed in the context of Fig.~\ref{Fig:DivP_Sym}, in the present case of two largely distinct pulses, one might have expected a different behavior at first sight. The temporal ordering seems to be very relevant here, since the dynamical phase $\varphi$ depends on the laser-dressed energy $E_L$ associated with the first pulse only [see Eq.~\eqref{Eq_DynPh}]. For example, when the high-intensity pulse B arrives first, the dynamical phase $\varphi$ accounts for the strong action of pulse B on the classical dynamics of the particles, which is described by the term $H_1^\star$ [see Eq.~\eqref{dynph_orig}]. Accordingly, the momentum dependence of $\varphi$ is substantially enhanced in comparison to the case when the weaker pulse A arrives first. One has to keep in mind, though, that the pair-creation amplitude $F_1$ of the first (strong) pulse involves the same Volkov phases $H_1$ [see Eqs.~\eqref{EqIj} -- \eqref{SDP_F}]. It thus exhibits a similar behavior as the dynamical phase factor $e^{−i\varphi}$. The latter only takes care that the quantum phase has properly evolved until the second pulse arrives. This property is independent of the pulse intensities and their sequence. Therefore, the invariance of the pair production probability under exchange of the pulses for $\chi_A = \chi_B = 0$ can be understood within the picture of two symmetrically formed slits, which was developed in the context of Fig.~\ref{Fig:DivP_Sym}.

\section{Conclusion}
The broad phenomenology of SFBW pair production in double laser pulses has been studied. We focused on the influence of a time delay $D$ between the pulses, which may equal or differ from each other. This way, previous investigations of SFBW pair production in short laser pulses have been extended, including a recent study which treats the process in a sequence of identical pulses following each other with $D=0$ \cite{Krajewska2014}.

We have shown that the presence of the second laser pulse induces characteristic interference effects, both in the energy spectra of produced particles and the total process probability. These effects are very sensitive to the time delay and particularly pronounced when the pulses are separated by relatively small distances on the order of the pulse lengths. In general, the total production probability exhibits (quasi-)periodic, damped oscillations when $D$ grows. For the case of two identical pulses, we were able to explain this functional dependence by a simplified analytical model. Note that similar features were observed in differential probabilities for pair production in time-dependent electric field pulses  \cite{Akkermans2012,Li2014,Kohlfurst2013}.
Regarding the two pulses as constituents of one combined field, the strong dependence on the pulse distance demonstrates the high sensitivity of the SFBW process on the properties of the driving laser field.

The latter conclusion is also in accordance with our finding that the temporal order of two non-identical pulses generally matters. The ordering can strongly influence the production probability for a given value of $D$ and even qualitatively modify the functional $D$-dependence. Only for special choices of the CEPs, the production probability was found invariant under exchange of the pulse sequence. The relation between the precise shapes of the pulses and the relevance of their ordering may be understood intuitively by interpreting the two consecutive laser pulses as a, possibly asymmetric, double-slit aperture which is simultaneously probed by the gamma quantum.

It is an interesting question to which extent a gamma beam of finite duration would modify the present results. In particular, if the gamma pulse was shorter than the duration of the combined laser field, qualitative changes can be expected. This aspect shall be addressed in future work. 

\section*{Acknowledgement}
We thank Matthias Dellweg for fruitful discussions.

\section*{References}


\end{document}